\documentclass{PoS}

\title{Towards a composite Higgs and a partially composite top quark}

\ShortTitle{Composite Higgs and partially composite top}

\author{\speaker{B. Svetitsky},$^a$ V. Ayyar,$^b$ T. DeGrand,$^c$ M. Golterman,$^d$ D. Hackett,$^{ce}$ W. Jay,$^f$ E. Neil,$^c$ and Y. Shamir$^a$ \\
        \llap{$^a$}Raymond and Beverly Sackler School of Physics and Astronomy, Tel Aviv University, 69978 Tel Aviv, Israel\\
        \llap{$^b$}NERSC, Lawrence Berkeley National Laboratory, Berkeley, California 94720, USA\\
        \llap{$^c$}Department of Physics, University of Colorado, Boulder, Colorado 80309, USA\\
        \llap{$^d$}Department of Physics and Astronomy, San Francisco State University, San Francisco, California 94132, USA\\
        \llap{$^e$}Center for Theoretical Physics, Massachusetts Institute of Technology, Cambridge, Massachusetts 02139, USA\\
        \llap{$^f$}Theoretical Physics Department, Fermi National Accelerator Laboratory, Batavia, Illinois 60510, USA}

\abstract{We have calculated quantities of interest to a theory of compositeness. The lattice model, approximating the candidate theory, is the SU(4) gauge theory coupled to fermions in two color representations. For the composite Higgs, a current correlator gives one of the ingredients of the effective Higgs potential. For the partially composite top quark, we have hyperbaryon matrix elements that govern mixing of the fundamental quark with its heavy composite partner. The matrix elements turn out to be so small that the theory is disfavored as a source of a realistic top mass.
}

\FullConference{37th International Symposium on Lattice Field Theory - Lattice2019\\
		16-22 June 2019\\
		Wuhan, China}
\def\LHC{\Lambda_{\textrm{\small HC}}}
\def\LEHC{\Lambda_{\textrm{\small EHC}}}
\def\LtHC{\Lambda_{\textrm{\tiny HC}}}
\def\LtEHC{\Lambda_{\textrm{\tiny EHC}}}
\def\gHC{g_{\textrm{\small HC}}}
\def\gEHC{g_{\textrm{\small EHC}}}

\begin{document}

\section{Introduction}
For some time we have been studying a lattice theory that is closely related to a model that goes beyond the Standard Model to generate a composite Higgs boson and a partially composite top quark.
I will present our most recent results \cite{Ayyar:2018glg,Ayyar:2019exp}, which are of direct relevance to the phenomenology of the theory but which are by no means the whole story.
Moreover, the original model has shortcomings; as we will see, some of our results make the original scenario even more difficult to realize.
Our work furnishes a prototype for such calculations in other models.

The phenomenological model \cite{Ferretti:2013kya,Ferretti:2014qta} is a ``Goldstone Higgs'' model \cite{Georgi:1984af,Dugan:1984hq}.
It is an SU(4) hypercolor gauge theory with a scale $\LHC$ that is perhaps 5~TeV\@.
There are 5 Majorana fermions~$Q$ in the sextet representation of the hypercolor SU(4)---the two-index antisymmetric representation.
When massless, these carry an SU(5) chiral symmetry, which breaks spontaneously to SO(5);
this SO(5) includes the gauge group of the electroweak theory as well as its custodial SU(2).
The Goldstone bosons include fields with the SM quantum numbers of the Higgs multiplet.
Thus the Higgs field $h$ has no mass and in fact no potential at all, which protects it from $\LHC$\@.
The Higgs potential $V(h)$, which is needed to create the Higgs {\em vev}, is to arise from loop diagrams involving the fermions and gauge bosons of the Standard Model.
While the gauge loop can only act to stabilize $v\equiv\langle h\rangle=0$, the top quark loop may (or may not) act to create a nonzero  $v$\@.
We have addressed the gauge loop contribution (see below).
We have no results as yet for the top loop contribution.

In addition to the sextet fermions $Q$, the theory contains three multiplets of Dirac fermions $q$ in the fundamental (quartet) rep of the hypercolor SU(4).
Their flavor SU(3) symmetry is gauged to become QCD\@.
Hypercolor confines the $q$'s and the $Q$'s to hypersinglet mesons and baryons.
Like the mesons, baryons made of $Q$'s alone or of $q$'s alone can only be bosons.
A novel {\em chimera} baryon, $Qqq$, is the only simple hypersinglet fermion in the hypercolor theory; it turns out to have the right SM quantum numbers to mix with a massless $t$ quark and give it a reasonable mass via a seesaw \cite{Kaplan:1991dc}.
This mixing is postulated to arise from four-fermi $tQqq$ interactions, a remnant of unknown interactions at some higher scale $\LEHC\gg\LHC$\@.

\section{The lattice theory}
Our lattice theory is a bit simplified from the above, for numerical convenience.
It is still an SU(4) gauge theory, but instead of 5 Majorana sextet fermions $Q$ we have 2 Dirac fermions (which amounts to 4 Majoranas).  Instead of 3 Dirac fermions $q$ in the quartet rep we have two.
The symmetry breaking scheme SU(4)$\to$SO(4) does {\em not\/} create enough Goldstone bosons to include the Higgs field, so this model is not phenomenologically adequate.
Still, for what we calculate, there are clear qualitative similarities to the model presented above.
We use Wilson--clover fermions for both multiplets, with nHYP smearing in the fermion actions and and an NDS gauge action~\cite{DeGrand:2014rwa}.

We have previously reported results for the spectrum of the theory.
We first calculated masses and decay constants for $0^-$ and $1^-$ mesons, both of the $\bar QQ$ and $\bar qq$ variety \cite{Ayyar:2017qdf}.
We then moved on to the baryons, including  $q^4$ and $Q^6$ baryons, all of which are bosons, and the more interesting $Qqq$ chimeras~\cite{Ayyar:2018zuk}.
In each of our ensembles we calculated the flow scale $t_0/a^2$ in order to fix its lattice spacing.
Then a global fit to two-representation chiral perturbation theory \cite{DeGrand:2016pgq} allowed extrapolation to a sensible chiral and continuum limit.

On to the new results.
I will present first our calculation of the matrix element that mixes the chimera baryon with the initially massless top quark \cite{Ayyar:2018glg}.
This is where a concrete calculation poses a new difficulty for this class of model.
Then I will turn to the composite Higgs potential, where we give only a partial calculation \cite {Ayyar:2019exp}, perhaps all that is possible at present.

\section{Mixing with the top quark}
\begin{figure}
   \includegraphics[width=.8\textwidth]{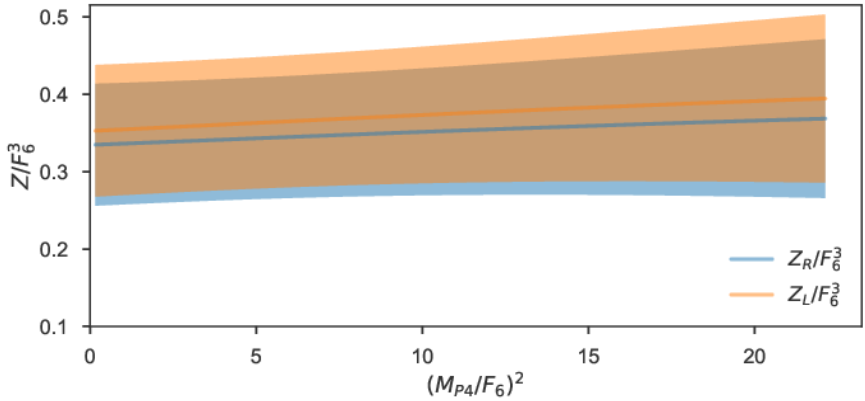}
     \caption{Chimera baryon matrix elements}
     \label{fig:Z}
     \end{figure}

The mixing interaction in the HC theory take the generic form
\begin{equation}
V_\textrm{\small mix} = G_R \bar{t}_L B_R + G_L \bar{t}_R B_L + \textrm{h.c.},
\end{equation}
where $t$ is the massless, fundamental top quark field and $B=Qqq$ is, schematically, the field of the chimera baryon whose mass is $M_B\sim\LHC$\@.
The effective couplings have their origins in the EHC sector as $G_{L,R}\sim \gEHC^2/\LEHC^2$\@.
The resulting mixing of the $t$ with the $B$ will give the physical top a mass,
\begin{equation}
m_t \approx	G_L G_R \frac{Z_L Z_R}{M_B} \frac{v}{F_6},
\label{eq:mt}
\end{equation}
where $F_6$ is the decay constant of $P_6$, the pseudoscalar meson $\bar QQ$ made of sextet fermions; it can be used instead of $\LHC$ to represent the HC scale.
From $m_t$ we can define the top Yukawa coupling via $m_t= y_tv$\@.
The constants $Z_{L,R}$ are matrix elements of the local chimera baryon operator between the vacuum and the physical chimera state,
\begin{equation}
\left\langle0\left|(Qqq)_{L,R}^\alpha\right|\textrm{Chimera}\right\rangle
=Z_{L,R}\,\,u^\alpha,
\end{equation}
where
$u^\alpha$ is a Dirac spinor at ${\bf p}=0$\@.
(These are analogous to the matrix elements needed for calculating proton decay.)
We are interested in the chiral limit $m_6\to0$ for the sextet fermions in order that the Goldstone Higgs be massless, as well as taking the continuum limit.
The quartet fermions are not necessarily massless; instead of their $m_4$ we take the mass of $P_4$, the $\bar qq$ pseudoscalar, as the $x$-axis for plots.
The result of our calculation, shown in Fig.~\ref{fig:Z}, is that the two $Z$'s are about equal and independent of the quartet mass.
     
The surprise is that the matrix elements are small,
\begin{equation}
\frac{Z}{F_6^3}=0.35(8),
\end{equation}
which we compare to QCD, where $Z/f_\pi^3\simeq7$\@.
The top mass $m_t$ and its Yukawa coupling $y_t$ are proportional to the square of this ratio;
inserting it into Eq.~(\ref{eq:mt}) along with our measured value for the ratio $M_B/F_6\simeq6$ gives
\begin{equation}
y_t \simeq 0.01	\left(\frac{\gEHC}{\LEHC}\,F_6\right)^4.
\label{eq:yt}
\end{equation}
We know that $y_t\simeq1$ and, whatever $\LEHC$ is, certainly $\gEHC<1$ there.
This gives us an upper bound $\LEHC<F_6/3$, which is a reversal of the assumed hierarchy $\LEHC\gg\LHC$\@.

To put it another way: If $\LEHC$ is made large enough that the EHC theory doesn't wreck low-energy physics, the top quark mixing and, hence, the top quark mass come out unrealistically small.

\section{Mixing: discussion \label{sec:disc}}
This points to a general problem with this type of model, that was recognized before our lattice result made it concrete (see \cite{Panico:2015jxa} for a discussion).
Even if, in some other theory, $Z/F_6^3$ were to come out much larger, it still limits how high you can push $\LEHC$ and still obtain a reasonable top mass, because of the fourth power in Eq.~(\ref{eq:yt}).
There is, however, a possible out that may emerge from running of the four-fermi couplings.
$G_{L,R}$ are defined as $\sim \gEHC^2/\LEHC^2$ at the EHC scale; they might run to larger values at $\LHC$ where we measure $Z$, according to
\begin{equation}
G(\LHC) =
G(\LEHC) \exp \left( -\int_{\LtHC}^{\LtEHC} \gamma_B(\gHC(\mu)) \frac{d\mu}{\mu} \right),
\end{equation}
where $\gamma_B$ is the anomalous dimension of the chimera baryon operator.
If $\gamma_B$ is large and negative, $G(\LHC)$ could receive a considerable enhancement.
Unfortunately, our theory is a conventional gauge theory with asymptotic freedom and confinement, with a small, perturbative anomalous dimension for the entire energy range~\cite{DeGrand:2015yna,Pica:2016rmv}.

One might follow a long-range strategy to look for other models with large anomalous dimension $\gamma_B$ {\em and\/} large matrix elements $Z$\@.
One approach resembles the search for walking technicolor, namely, to try to approach the conformal window from the present model by adding fermion flavors.
These would slow the running of the gauge coupling; moreover, 
a large anomalous dimension might appear near the conformal window. 
One could also look at other models entirely, among those listed by Franzosi and Ferretti \cite{BuarqueFranzosi:2019eee}.
Another theory that is under current study \cite{Bennett:2017kga,Bennett:2019jzz,Bennett:2019ckt} is an Sp(4) gauge theory with a global SU(4) symmetry that breaks to Sp(4). 
See also Ref.~\cite{Gertov:2019yqo}.

\section{Composite Higgs potential}

The effective potential for the Goldstone fields in a model like ours can be quite complex, since there are usually more fields than are needed for the SM Higgs multiplet.
If one considers only the latter, however, the generic form of the potential can be written as
\begin{equation}
V_{\rm eff}(h)=-\alpha\cos^2(h/f)-\beta\sin^2(2h/f).
\label{eq:Veff}
\end{equation}
The first term stabilizes the $h=0$ vacuum if $\alpha>0$ and hence prevents the SM Higgs mechanism.
$\alpha$ contains a positive piece due to the SM gauge fields as well as pieces from top loops, not necessarily positive:
\begin{equation}
\alpha=\frac12(3g^2+g^{\prime2})C_{LR}\,+\, \textrm{top loops}.
\end{equation}
The second term in Eq.~(\ref{eq:Veff}) comes from top loops alone.

A calculation of $C_{LR}$ is a beginning.
It comes from a current--current correlation function,
\begin{equation}
  C_{LR} = \int_0^\infty dq^2 q^2\, \Pi_{LR}(q^2),
  \label{eq:CLR}
\end{equation}
where
\begin{equation}
(q^2 \delta_{\mu\nu}-q_\mu q_\nu)\, \Pi_{LR}(q^2) \rule{0ex}{3ex}
  = -\int d^4x\, e^{iqx} \left\langle J_\mu^L(x) J_\nu^R(0)\right\rangle.
\end{equation}
Here $J^{L,R}$ are the chiral currents of the sextet fermions.

\begin{figure}
    \begin{center} 
   \includegraphics[width=.45\textwidth]{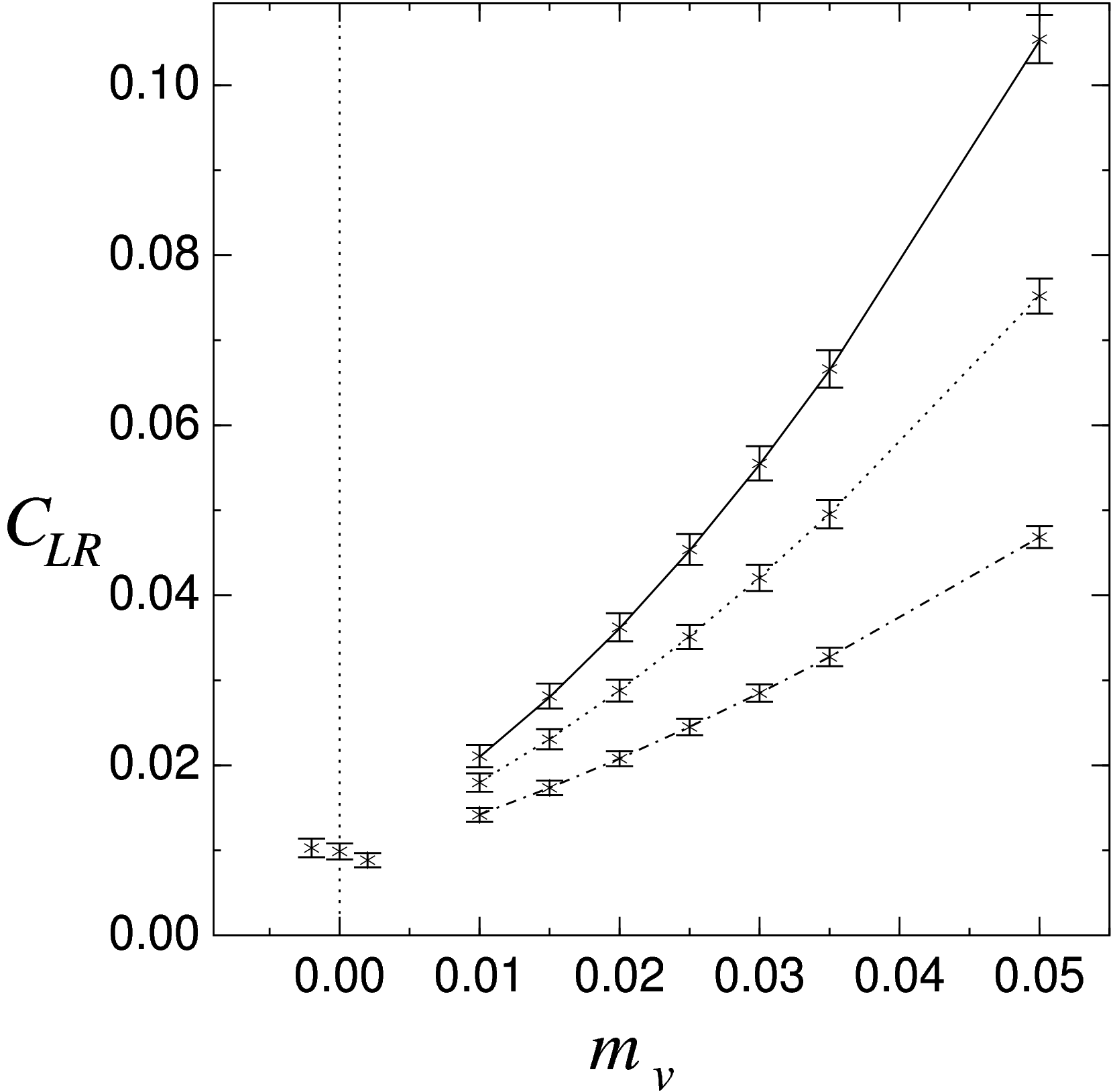}
    \end{center}
     \caption{Gauge term in the Higgs effective potential, as a function of valence mass $m_v$, in a single ensemble.
     The extrapolations to $m_v=0$ are displaced horizontally for clarity.}
     \label{fig:CLR}
     \end{figure}

We have previously calculated $C_{LR}$ in a restricted model, containing sextet fermions but no quartet fermons \cite{DeGrand:2016htl}.
We have now added the quartet fermions, and also simplified the calculation by defining the valence currents with smeared staggered fermions rather than overlap fermions.
Exact chiral symmetry is needed to force a severe UV divergence to vanish in the valence chiral limit.
An illustration of how this works out is given in Fig.~\ref{fig:CLR}, showing $C_{LR}(m_v)$ in a single ensemble as a function of the valence fermion mass $m_v$\@.
The three curves represent results of integrating Eq.~(\ref{eq:CLR}) with three different UV cutoffs, and we see that the extrapolation $m_v\to0$ gives a cutoff-independent result.

Determination of the flow scale and the current fermion masses makes possible an extrapolation to chiral and continuum limits.
As for the top quark mixing amplitude presented above, we are interested in the chiral limit in the (valence and dynamical) sextet fermions.
Again, we find no discernible dependence on the quartet mass $m_4$, within admittedly large error bars.
The result is
\begin{equation}
\frac{C_{LR}}{F_6^4}=29(8)(5).\label{eq:CLRresult}
\end{equation}
The counterpart of $C_{LR}$ in QCD is related to the electromagnetic mass splitting of the pions \cite{Das:1967it},
\begin{equation}
m_{\pi^\pm}^2 - m_{\pi^0}^2
= \frac{3\alpha}{4\pi} \frac{C_{LR}}{f_\pi^2}.
\end{equation} 
This gives $C_{LR}/f_{\pi}^4\approx 42$ in QCD, which is not too different from Eq.~(\ref{eq:CLRresult}).

\section{Outlook}

From our result for $C_{LR}$ one might hope that QCD is a good basis for guesswork in models such as ours.
This conclusion is vitiated, however, by what we find for the mixing matrix elements $Z_{L,R}$, which are quite different from their analogue in QCD\@.
In the end, there is no substitute for actually calculating something.
Of course, this applies just as strongly to the Higgs effective potential, where we have calculated only the gauge loop contribution.
In the absence of a complete calculation of the top quark loop terms, there is no proof that the Goldstone Higgs paradigm works at all.

A complete calculation of the Higgs potential is a tall order.
It might even be pointless.
The difficulties are both formal and phenomenological.
Formally, an exact calculation would require extraction of four-point functions of the chimera operator and their integration in momentum space.
This can be short-circuited by saturating the four-point function with single-particle intermediate states, with turns the problem into a product of two-point functions.
[We did something similar in writing Eq.~(\ref{eq:mt}): saturating the chimera propagator with the single-chimera state.]
Organizing the effective potential with chiral perturbation theory still leaves a good number of low-energy constants to calculate \cite{Golterman:2015zwa,Golterman:2017vdj}.
This is only the beginning, however.
The final expression will still depend on four-fermi couplings inherited from the EHC scale; there are many of them and there is no guiding principle to winnow them down.

On top of this, one still has to find a HC theory with a large anomalous dimension for the chimera operator, for reasons made clear in Sec.~\ref{sec:disc}.

\end{document}